\documentclass[12pt]{article}
\usepackage{amsmath}
\usepackage{graphicx}
\usepackage{enumerate}
\usepackage[numbers]{natbib}
\usepackage{url} % not crucial - just used below for the URL 
\usepackage[colorlinks=true,linkcolor=blue,citecolor=blue,urlcolor=blue,]{hyperref}
\usepackage{amsfonts}
\usepackage{amsmath}
\usepackage{amsthm}
\usepackage{booktabs}
\usepackage{steinmetz}
\usepackage{pdfpages}
\usepackage{tikz}
\usepackage{eso-pic}
\usepackage{fancyhdr}
\usepackage{algorithm}
\usepackage{algorithmic}
\DeclareGraphicsExtensions{.eps,.ps,.jpg,.bmp}

\newtheorem{theorem}{Theorem}

\newtheorem{proposition}[theorem]{Proposition}

\newtheorem*{mexample}{Motivation Example}

\newcommand{\blind}{1}

\begin{document}

\def\spacingset#1{\renewcommand{\baselinestretch}%
{#1}\small\normalsize} \spacingset{1}

%%%%%%%%%%%%%%%%%%%%%%%%%%%%%%%%%%%%%%%%%%%%%%%%%%%%%%%%%%%%%%%%%%%%%%%%%%%%%%

\if1\blind
{
  \title{\bf Zero-inflated Smoothing Spline (ZISS) Models for Individual-level Single-cell Temporal Data}
  \author{Yifu Tang\\
  Tsinghua University\\
\texttt{tang-yf20@mails.tsinghua.edu.cn}
    \and 
    Yi Zhang\\
    University of Arizona\\
    \texttt{yizhang6@arizona.edu
}
\and
    Yue Wang\\
    University of Colorado\\
    \texttt{yue.2.wang@cuanschutz.edu}
    \and
    Jingyi Zhang*\\
    Tsinghua University\\
    \texttt{jingyizhang@mail.tsinghua.edu.cn}
    \and 
        Xiaoxiao Sun* \\
    University of Arizona\\
    \texttt{xiaosun@arizona.edu}
    }
  \maketitle
} \fi

\if0\blind
{
  \bigskip
  \bigskip
  \bigskip
  \begin{center}
    {\LARGE\bf Zero-inflated Smoothing Spline (ZISS) Models for Individual-level Single-cell Temporal Data}
\end{center}
  \medskip
} \fi

\bigskip
\begin{abstract}
Recent advancements in single-cell RNA-sequencing (scRNA-seq) have enhanced our understanding of cell heterogeneity at a high resolution. With the ability to sequence over 10,000 cells per hour, researchers can collect large scRNA-seq datasets for different participants, offering an opportunity to study the temporal progression of individual-level single-cell data. However, the presence of excessive zeros, a common issue in scRNA-seq, significantly impacts regression/association analysis, potentially leading to biased estimates in downstream analysis. Addressing these challenges, we introduce the Zero Inflated Smoothing Spline (ZISS) method, specifically designed to model single-cell temporal data. The ZISS method encompasses two components for modeling gene expression patterns over time and handling excessive zeros. Our approach employs the smoothing spline ANOVA model, providing robust estimates of mean functions and zero probabilities for irregularly observed single-cell temporal data compared to existing methods in our simulation studies and real data analysis. 
\end{abstract}

\noindent%
\vfill

* To whom correspondence should be addressed.

\newpage
\spacingset{1.9} % DON'T change the spacing!
\section{Introduction}
\label{sec:intro}

Recent advances in single-cell RNA-sequencing (scRNA-seq) techniques hold great promise in improving our understanding of cell-to-cell heterogeneity and cell differentiation at a high resolution \cite{wu2014quantitative,buettner2015computational, papalexi2018single}. With the aid of high-throughput sequencing methods, researchers can process over 10,000 cells per hour, enabling the collection of large scRNA-seq datasets from multiple subjects with different disease statuses \cite{aviv2017human}. In the realm of single-cell data analysis, trajectory inference methods stand out as pivotal tools. Their main goal is to order individual cells along a trajectory, often referred to as pseudotime, based on their gene expression profiles, resulting in the generation of single-cell temporal data at the subject level. \cite{saelens2019comparison,cannoodt2016computational,liu2017reconstructing}. This approach opens up fresh avenues for studying biological questions at the high-resolution level, including association studies between the phenotypes of participants and cellular processes over time. These investigations play a crucial role in unraveling the complexities of disease progression and are thus of utmost significance in scientific research \cite{tanay2017scaling,etzrodt2014quantitative}.

Due to the low capturing and sequencing efficiency in scRNA-seq, there is a common dropout issue, referring to the presence of excessive zero counts in the single-cell data \cite{chen2016single,li2018accurate,qiu2020embracing}. This zero-inflation phenomenon notably impacts data analysis, potentially skewing the inferred patterns of true associations of interest. In addition, for the single-cell temporal data, zero proportions are changing over pseudotime. Gene expression data are also correlated across pseudotime. Disregarding these features of single-cell temporal data during data analysis can lead to estimation bias, as demonstrated in the following motivation example. 

\begin{mexample}
In this motivation example, we analyzed the scRNA-seq data obtained from a subject with acute kidney failure \cite{lake2021atlas}. Pseudotime estimation was obtained for the data \cite{ji2016tscan, hou2021statistical}. Our objective is to estimate the smooth gene expression curve over time from the zero-inflated noisy data. The resultant curve plays a crucial role in downstream analysis of scRNA-seq data, such as differential analysis. In Figure~\ref{fig:motive1} a), we show the curves fitted for a subject with acute kidney failure using the proposed method (ZISS) and the original smoothing spline method applied to the data without zeros (NZSS) \cite{gu2013smoothing}. The original smoothing spline method did not converge when applied to the original data due to the excessive zeros. Hence, it is crucial to address the issue of excessive zeros in modeling single-cell temporal data. In Figure~\ref{fig:motive1} b), we present the data distribution across pseudotime for the same gene discussed in a). This illustration shows that the proportions of zeros vary throughout the pseudotime, and the number of observations also differs at each pseudotime point. Moreover, the proportions of zeros exhibit correlation over pseudotime. These factors must be taken into account when estimating the probabilities of zeros.
\begin{figure}[h!]
%\centering
\includegraphics[height=5.1cm]{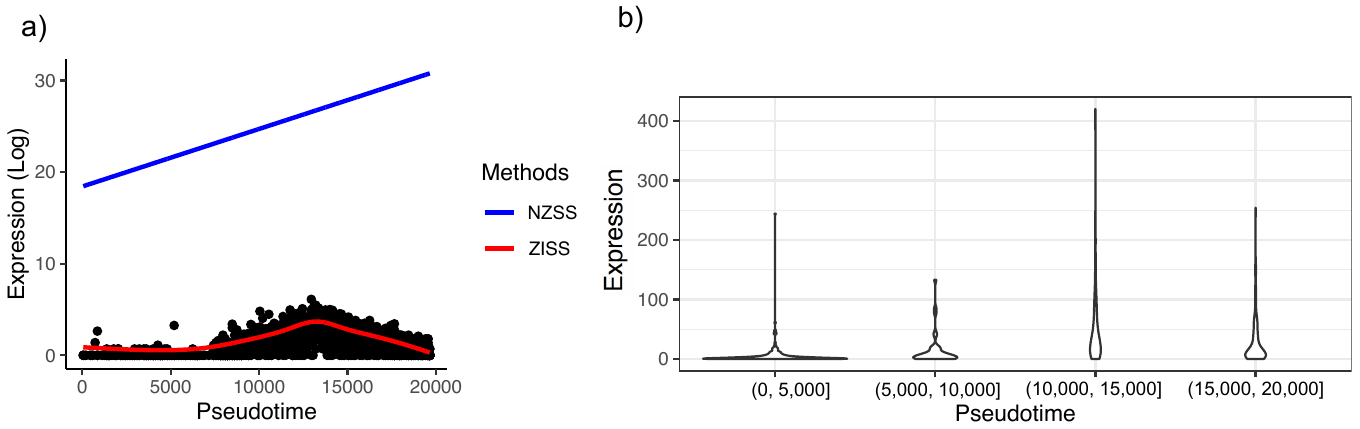}
\caption{The motivation example. a). The black dots represent gene expression levels along pseudotime for different cells from a gene (Ensembl ID: ENSG00000137673).  The red curve represents the estimated expression profiles using the proposed method (ZISS), while the blue curve represents the estimated curve obtained by applying the original smoothing spline method to the data without zeros (NZSS). b). The violin plots depict gene expression across pseudotime, which has been divided into four distinct intervals for analysis.}
\end{figure}\label{fig:motive1}
\end{mexample}

In analyzing single-cell temporal data, the methods developed must tackle two significant challenges, as illustrated in the motivation example. First, addressing the issue of excessive zeros is imperative to avoid numerical errors and estimation bias. Second, considering the correlations between cells and their corresponding gene expressions over time, it becomes essential to incorporate smoothness assumptions when estimating the temporal patterns of gene expression levels from zero-inflated data. To tackle the first challenge, a commonly employed approach involves constructing a two-component mixture model. This model specifies the distribution of a random variable by using a probabilistic mixture of zero and a regular component, the distribution of which typically belongs to the exponential family. Several approaches have been developed for the analysis of non-temporal scRNA-seq data \cite{jiang2022statistics, fang2016zero, lord2005poisson}. These methods effectively tackle the challenges posed by excessive zeros using zero-inflated Poisson and/or negative binomial models. However, these methods are only applicable to regular scRNA-seq data and are not suitable for the analysis of single-cell temporal data. To surmount the second challenge, semiparametric models were considered to analyze zero-inflated temporal data \cite{fahrmeir2006structured, xue2009bayesian}. These models involve modeling the non-zero-inflated response using flexible semiparametric predictors, while the probability of zero-inflation is constrained to an unknown constant. However, in practical applications, the assumption that the zero-inflation probability remains constant over time may not hold. To address this, zero-inflated generalized additive models (ZIGAMs) were developed to extend the models to more general cases \cite{barry2002generalized}. ZIGAM employs smooth functions to model the mean pattern of the non-zero-inflated response along with time and the non-zero-inflation probability, then estimates the two smooth functions through two GAMs. Building on ZIGAM, the Constraint ZIGAM, or the COZIGAM, \cite{liu2011generalized, liu2008constrained, liu2010introducing} was introduced to establish a connection between the mean function of the response and the non-zero-inflation function in cases where both processes may be influenced by some common factors. To estimate the unknown link functions, ZIGAM and COZIGAM utilize the regression spline method. However, these methods may encounter challenges when pseudotime points are irregularly spaced or records are unevenly distributed along pseudotime, as the regression spline estimation is highly sensitive to the choice of knots and can prove to be unreliable in the presence of varying levels of sparsity. Therefore, novel approaches are needed to analyze the single-cell temporal data. Additionally, the single-cell temporal data exhibits varying numbers of observations over time, posing numerical challenges for estimation using regression spline methods. Our approach, utilizing smoothing spline ANOVA models, effectively addresses this issue.

%When analyzing temporal zero-inflated data, semiparametric models were considered \cite{fahrmeir2006structured, xue2009bayesian}. In the semiparametric models, the non-zero-inflated response was modeled by flexible semiparametric predictors, and the probability of zero-inflation was restricted to an unknown constant. In many real-world applications, the assumption that the zero-inflation probability remains unchanged over time may not hold. Zero-inflated generalized additive models (ZIGAMs) have been developed to extend the models to more general cases \cite{barry2002generalized}. ZIGAM uses smooth functions to describe the mean pattern of the non-zero-inflated response along with time and the non-zero-inflation probability, then estimates the two smooth functions through two GAMs. Based on ZIGAM, COZIGAM \cite{liu2011generalized} was developed to link the function of the mean pattern to the function of the non-zero-inflation for cases where the two processes may be affected by some common factors. To estimate the unknown link functions, ZIGAM and COZIGAM apply the regression spline method, which may be problematic when the pseudotime points are not equally-spaced, or the records are not uniformly distributed along pseudotime. Under such cases, the regression spline estimation is extremely sensitive to the choice of knots and always turns out to be unrobust when various levels of sparsity exist. 

In this paper, we introduce the ZISS method for the analysis of single-cell temporal data. Our approach is specifically designed to address the challenges posed by excessive zeros in scRNA-seq data, stemming from both biological and technical factors. While biological zeros represent genuine absence of gene expression, technical zeros arise due to limited genetic materials in individual cells. Our proposed model comprises two components: one for modeling the underlying gene expression patterns over time and another for accommodating the excessive zeros. We utilize the smoothing spline ANOVA model to estimate the link functions, offering greater robustness across various sparsity levels by mitigating the dependence on knot selection. Furthermore, our method yields more precise estimates of link functions and zero probabilities compared to existing approaches.

%\begin{figure}[h!]
%\centering
%\includegraphics[height=6.5cm]{Violin.pdf}
%\caption{Visualization of data within different pseudo time periods \textcolor{red}{need to be modified to the real data's pattern}}
%\end{figure}

The remainder of the article is structured as follows. Section 2 presents the model setup for the proposed ZISS model, along with an overview of the model estimation process. In Sections 3 and 4, we present the results of simulation studies and real data analysis, respectively. Section 5 provides additional remarks and conclusions. Notations and derivation are provided in the Appendix for reference.

\section{Model Formulation and Estimation}
In this section, we present a novel Zero Inflated Smoothing Spline (ZISS) method and provide estimation details of this approach. A two-component mixture model will be employed to define the data distribution. One component will be used to model the true signals based on the smoothing spline ANOVA models, whereas the other component will be utilized for modeling the excessive zeros. To adaptively manage the fluctuating zero probabilities over time, we suggest the utilization of B-spline basis functions to model the excessive zero probability functions. The estimation of the model parameters will be conducted using the Expectation–Maximization (EM) algorithm. 
The Newton-Raphson algorithm is implemented to estimate the zero probability functions.  
\subsection{Zero Inflated Smoothing Spline (ZISS)}
As demonstrated in the motivation example (see Figure~\ref{fig:motive1}), the temporal dependence should be considered not only in the underlying non-zero inflated gene expression levels, but also in the probability of excessive zeros when analyzing single-cell temporal data. To formulate the temporal dependence, we denote the underlying non-zero inflated gene expression level as $\mu(t)$, and the probability of zero-inflation, which is also referred to as the dropout rate, as $p(t)$. 
For each gene, we further assume the gene expression data are observed at $N$ pseudotime points, $t_{min}< t_1 < t_2 <\cdots<t_N< t_{max}$. 
%For a gene, without loss of generality, we assume that the pseudo-time interval is $[0,1]$. There are $N$ time points, $0\leq t_1 < t_2 <\cdots<t_N\leq 1$, in the pseudo-time interval (not necessarily uniform), 
At each pseudotime point $t_i$, we observe $M_{i}$ cells, which are always recorded as the counting responses $y_{i,1},\cdots,y_{i,M_{i}} \in \mathbb{N}$, where $\mathbb{N}$ denotes the set of non-negative integers. 
We then consider the zero-inflated model as follows, 
% As a result, the obtained count data, coupled with the corresponding pseudo-time, is denoted as
% $$\left\{t_i, y_{i,j}\right\}_{1\leq i \leq N,1\leq j \leq M_{i}}$$
% with latent variables $g_{i,j}$ indicating the distribution of $y_{i,j}$, i.e.
\begin{equation}
\label{model}
 y_{i,j}\sim\begin{cases}
     \textbf{Poisson}(\mu(t_i)), \quad \textit{if}~g_{i,j} = 1,\\
     0, \quad \textit{if}~g_{i,j} = 0.
 \end{cases}
 \end{equation}
% For smooth function $\mu(t)$  and piece-wise smooth function $p(t)$ where $t$ denotes the time, the response with probability $p(t)$ having Poisson Distribution with parameter $\mu(t)$ and with probability $1-p(t)$ being 0.  It could be expressed as the following hierarchical model: At pseudo-time $t$, 
%\begin{equation}
%\label{model}
%y_{i,j} = g_{i,j} z_{i} + \left(1 - g_{i,j}\right)\delta_0, 
%\end{equation}
where $g_{i,j}\sim Ber(p(t_i))$ is a Bernoulli variable indicating zero-inflation. The zero probability function $p(t_i)$ is used to model the excessive zeros. When $g_{i,j}=1$, $y_{i,j}$ follows a Poisson distribution with a mean of $\mu(t_i)$ at pseudotime $t_i$, and is zero otherwise. %; when $g_{i,j}=0$, $y_{i,j} = \delta_0$, the Dirac measure at 0. 

To address the irregularly observed single-cell temporal data, we propose utilizing the smoothing spline ANOVA model \cite{gu2013smoothing} to estimate the mean function $\mu$. This model is knot-free, making it suitable for irregularly observed data. In particular, we consider the mean function in a reproducing kernel Hilbert space, 
$\mathcal{H} = \left\{ \mu: J(\mu) < \infty\right\},  $
where $J(\mu)$ is a quadratic functional \cite[Section 2.4.5]{gu2013smoothing}, measuring the roughness of functions. For example, for a twice differentiable function $\eta :[0,1]\rightarrow \mathbb{R}$, $J(\cdot)$ can be chosen as 
$J(\eta) = \int_{0}^1 \left(\frac{d^2\eta}{dx^2} \right)^2 dx.$
We decompose the space $\mathcal{H}$ into the null (kernel) space of $J(\cdot)$ and its orthogonal complement $\mathcal{H}_J$, i.e. $\mathcal{H} = \mathcal{N}_{J} \bigoplus \mathcal{H}_J $, where the null space of $J$ is defined as $\mathcal{N}_{J} = \left\{ \mu: J(\mu) = 0\right\}.$
%From the decomposition, $\mathcal{H}_J$ is an RKHS with norm $J()$. 
%Consider a quadratic functional $J(\cdot)$ which measures the roughness of some function, where $J(\mu) = J(\mu,\mu)$. We consider a reproducing kernel Hilbert space 
Based on the Representer Theorem \cite{gu2013smoothing}, the mean function $\mu(t)$ has the following form,
%\cite{gu2013smoothing} suggests that the minimizer of (Equ.\ref{pen}) $\mu_{\lambda}$ has the form 
\begin{equation}
\label{eq:representer}
\mu(t) = \sum_{\nu=1}^m d_{\nu} \phi_{\nu}(t) + \sum_{i=1}^n c_i R_J(t_i,t) = \boldsymbol{\phi} ^{\top} \boldsymbol{d} + \boldsymbol{\xi}^{\top} \boldsymbol{c},
\end{equation}
where $\left\{ \phi_{\nu} \right\}_{\nu=1}^m$ are basis functions in $\mathcal{N}_J$, $R_J(\cdot,\cdot)$ is the kernel function of $\mathcal{H}_J$, $\boldsymbol{\xi}$ and $\boldsymbol{\phi}$ are the vectors of the corresponding basis functions with coefficient vectors $\boldsymbol{c}$ and $\boldsymbol{d}$. For the Gaussian-type responses, the penalized least squares functional in a reproducing kernel Hilbert space $\mathcal{H}$ can be written as $L(\mu) + (\lambda/2) J(\mu)$, where $L(\mu)$ measures the goodness-of-fit between the responses and the function $\mu$ (e.g., least squares) and $\lambda$ is the smoothing parameter to control the smoothness of the function estimate. We then plug the equation~\eqref{eq:representer} into the penalized least squares functional to obtain the estimates of coefficients. More details about the estimation for Gaussian-type responses can be found in \cite[Chapter 3]{gu2013smoothing}. In line with the Gaussian cases, we will apply the penalized likelihood method to estimate the mean function $\mu$ in the Poisson processes. %Suppose the vector $\boldsymbol{Y}$ denotes the response observations, while ${S} = (\phi_j(t_i))_{1\leq i \leq n,1\leq j \leq m}$, ${R} = (R_J(t_i,t_j))_{i,j}$, then the coefficients $\boldsymbol{c}$ and $\boldsymbol{d} $ is obtained by 
%$$(\boldsymbol{c},\boldsymbol{d}) = \arg\min_{\boldsymbol{c},\boldsymbol{d}} = \frac{1}{n} \| \boldsymbol{Y} - {S}\boldsymbol{d} - {R}\boldsymbol{c}\|^2 + \lambda \boldsymbol{c}^{\top} {R} \boldsymbol{c}.$$
%Readers may see \cite[sections 1,2,3, and 5]{gu2013smoothing} for more details.

In the motivation example, the zero probabilities exhibit temporal dependency. Therefore, it is reasonable to assume that the probability is a smooth function. We assume that $p(t)$ is a piece-wise smooth function, with a summation equal to one over the temporal domain. Specifically, we assume that $$p(t) = \frac{1}{1+\exp(\sum_{i=1}^m\alpha_i b_i(t))},$$
where $b_i(t)$ are the B-spline basis functions over the pseudotime interval and $\alpha_1,\cdots,\alpha_m\in \mathbb{R}$ are the corresponding coefficients of these basis functions. This nonparametric form offers a flexible and smooth estimation of zero probabilities over time. In addition, it simplify the overall estimation process.

%Newton iteration is applied to solve the problem $(P_1)$. 

%Namely, at the time $t_i$, we define a latent Bernoulli variable $g_{i,j}\sim Ber(p(t_i))$ for the $j$th response, i.e. $g_{i,j}=1$ with probability $p(t_i)$. If $g_{i,j}=1$, then $y_{i,j}$ is generated by $Poisson(\mu(t_i))$. If $g_{i,j}=0$, then $y_{i,j}=0$. %Without ambiguity, we denote $y_{t_i,j} = y_{i,j}$ and $g_{t_i,j} = g_{i,j}$.

%Integrating this temporal dependency and applying smoothness assumptions to both the non-zero probability function $p(t)$ and gene expression curves $\mu(t)$ over time becomes an intuitive step. 

%In order to achieve these goals, we adopt a penalized likelihood approach to estimate the smooth function $\mu(t)$, utilizing B-spline basis functions for the accurate approximation of the continuous zero probability function $p(t)$. This approach enables us to effectively capture the temporal dynamics of zero proportions as well as transcriptional dynamics over time.

%Now that we have completed the definition of the model, 

\subsection{Estimation}
We utilize the maximum penalized likelihood method to estimate the mean function $\mu(t)$ and the probability function $p(t)$. In particular, given observations $t_i,y_{i,j}$, the latent variables $g_{i,j}$, and the roughness functional $J(\cdot)$, we minimize the following penalized negative log-likelihood functional to obtain the estimates of $\mu(t)$ and $p(t)$,
\begin{equation}\label{likelihood}
\begin{aligned}l(p(\cdot), \mu(\cdot); g_{i,j}, y_{i,j}) = &-\sum_{i=1}^N \sum_{j=1}^{M_{i}} \left(g_{i,j} \log p(t_i) + (1-g_{i,j}) \log(1-p(t_i)) \right) \\
&- \sum_{i=1}^N \sum_{j=1}^{M_{i}} \left( g_{i,j} y_{i,j} \log \mu(t_i) -g_{i,j} \mu(t_i) \right) + \frac{\lambda}{2} J(\mu) \\
& + \sum_{i=1}^N \sum_{j=1}^{M_{i}} g_{i,j}\log(y_{i,j}!),
\end{aligned}
\end{equation}
where the smoothing parameter $\lambda$ controls the trade-off between the goodness-of-fit of the model and the roughness of $\mu(t)$. As the latent variables are not observed, we utilize the Expectation-Maximization (EM) algorithm to perform estimation \cite{moon1996expectation}. In the Expectation (E) step, we compute the conditional expectation of the latent variable given the data and the current estimates of the mean and probability functions.  In the Maximization (M) step, we maximize the penalized likelihood functional to obtain the estimates for the mean and probability functions. In the estimation of $\mu(t)$ and $p(t)$, the last term of the penalized negative likelihood, as shown in~\eqref{likelihood}, is not involved. Therefore, it is excluded during the EM process. As a result, the M step is further divided into two sub-problems, both optimized simultaneously. For the first problem, we employ the Newton-Raphson method to obtain the estimate of the probability function. For the second problem, we utilize the penalized likelihood method to obtain estimates of the mean function. In the following, we will present the EM steps in estimation. The derivation details can be found in Appendix.

%They will be introduced in the following part of this section, and the constant $C$ only depends on the responses $y_{i,j}$ and latent variables $g_{i,j}$, which will not be included in our updating process of functions $\mu$ and $p$. 
%Note that in the real process,  the latent variable $g_{i,j}$ is always unobserved, we thus apply the Expectation-Maximization (EM) Algorithm \cite{xxx} to estimate $p(\cdot)$ and $\mu(\cdot)$ iteratively.

\paragraph{Expectation (E) step.} Denoting the fitted mean and probability functions as $\hat{\mu}_k$ and $\hat{p}_k$ at the $k$-th step (If $k=0$, $\hat{\mu}_k$ and $\hat{p}_k$ represent the initial values), we calculate the conditional expectation of the latent variable given the current estimates and observed data, as
$\mathbb{E}\left[ g_{i,j} | y_{i,j},\hat{\mu}_k,\hat{p}_k  \right]$ via the following proposition.
\begin{proposition}[Computing conditional expectation in the E step]
\label{cond-exp}
Given the observed data $\left\{t_i, y_{i,j}\right\}_{1\leq i \leq N,1\leq j \leq M_{i}}$ and the fitted curves $\hat{\mu}_k$ and $\hat{p}_k$ at the $k$-th ($k\ge 0$) step,  we have the conditional expectation of latent variable $g_{i,j}$,
\begin{equation}
\label{eq:conde}
\mathbb{E}\left[ g_{i,j} | y_{i,j}=0,\hat{\mu}_k,\hat{p}_k \right] = \frac{e^{-\hat{\mu}_k(t_i)} \hat{p}_k(t_i) }{e^{-\hat{\mu}_k(t_i)} \hat{p}_k(t_i) + 1-\hat{p}_k(t_i)},
\end{equation}
and 
$$\mathbb{E}\left[ g_{i,j} | y_{i,j}>0,\hat{\mu}_k,\hat{p}_k \right] = 1.$$
\end{proposition}

Denoting the obtained conditional expectation by $q_{i,j}=\mathbb{E}\left[ g_{i,j} | y_{i,j},\hat{\mu}_k,\hat{p}_k \right]$, we use the conditional expectation to replace the latent variables in the log-likelihood functional $l$ in the M step.

\paragraph{Maximization (M) step.} In the M step, we maximize the penalized log-likelihood functional given the conditional expectation in the previous E step. Equivalently, we minimize the negative penalized log-likelihood functional~\eqref{likelihood} via the following proposition. 
\begin{proposition}[Estimating $\mu$ and $p$ in the M Step]
    Given $q_{i,j}=\mathbb{E}\left[ g_{i,j} | y_{i,j},\hat{\mu}_k,\hat{p}_k \right]$ obtained in the E Step, we perform the following two optimization problems in the M step at the same time, %\textcolor{red}{($g_{i,j}$ or $q_{i,j}$ in the following equations? it should be $q_{i,j}$, I have modified.)}

    $$(P_1)~~\min_{p\in \mathcal{P}} -\sum_{i=1}^N \sum_{j=1}^{M_{i}} \left(q_{i,j} \log p(t_i) + (1-q_{i,j}) \log(1-p(t_i)) \right),$$
and
$$(P_2)~~\min_{\mu \in \mathcal{H}} - \sum_{i=1}^N \sum_{j=1}^{M_{i}} \left( q_{i,j} y_{i,j} \log \mu(t_i) -q_{i,j} \mu(t_i) \right) + \frac{\lambda}{2} J(\mu),$$
 where for given B-spline functions $b_1(t),\cdots,b_m(t)$ defined in the pseudo-time interval, $$\mathcal{P} = \left\{ \frac{1}{1+\exp(\sum_{l=1}^m\alpha_l b_l(t))}: \alpha_1,\cdots,\alpha_m \in \mathbb{R}\right\},$$ and $\mathcal{H}$ denotes the reproducing kernel Hilbert space wherein we confine the estimates of $\mu$.
\end{proposition}

For the zero probability function, we have $\log \frac{p(t)}{1-p(t)} = -\sum_{i=1}^m \alpha_i b_i(t)$. Thus in Problem $(P_1)$, the equation takes the form,
\begin{equation*}
\label{form-of-p}
q_{i,j}\log p(t) + (1-q_{i,j}) \log (1-p(t)) = -q_{i,j}\sum_{i=1}^m \alpha_i b_i(t) + \log(1-p(t)).
\end{equation*}
\begin{algorithm}[H]
\label{alg1}
        \caption{Zero inflated smoothing spline (ZISS)}
        \begin{algorithmic}[1]
        \STATE  \textbf{Input}: Smoothing parameter $\lambda$, single-cell data $\left\{ ( t_i,y_{i,j})\right\}_{1\leq i\leq N, 1\leq j\leq M_{i}}$, initial values $\hat{p}^{(0)}, \hat{\mu}^{(0)}$, converge criterion (threshold gap $\epsilon$) \\
        \FOR{$m$ in $1:$max-iteration step}

           \STATE (E Step) Compute the conditional expectation of $g_{i,n}$ to obtain $q_{i,j}^{(m)}$ from equation~\eqref{eq:conde}.
           
           \STATE (M Step) Minimize the objective functional to obtain $\hat{p}^{(m+1)}$ and $\hat{\mu}^{(m+1)}$ in $(P_1)$ and $(P_2)$
        \IF{converge criterion}
        \STATE Break
        \ENDIF

\ENDFOR
\IF{GCV in the final step}
        \STATE  Use the smoothing parameter $\lambda_0$ selected by GCV to fit final $\hat{\mu}$ and $\hat{p}$
\ENDIF
        \RETURN $\hat{p},\hat{\mu}$
        \end{algorithmic}
        \end{algorithm}

This formulation streamlines the minimization process. The second-order derivative of the first term with respect to $\alpha$ becomes zero, markedly reducing computational complexity in Newton's iteration. Moreover, it is essential that the probability function $p(t)$ remains within the [0,1] range. The chosen form of $p$ naturally ensures this constraint. In contrast, using a traditional B-spline function representation, such as $p(t) = \sum_{i}\alpha ' b_i(t)$, could result in values exceeding 1. 
%In the M-Step, we maximize the objective function by the method mentioned in \cite{gu2013smoothing}.
%We propose the stopping criterion as follows: The norm of the difference between $\hat{\mu}_n$ and $\hat{\mu}_{n-1}$ is smaller than a pre-set value or reaches the maximum iteration steps.
For each iteration, assuming the total number of samples is $n=N*M_i$, the theoretically optimal convergence rate for $\lambda$ is $O(n^{-\frac{2}{9}})$ \cite[Section 9]{gu2013smoothing}. In our algorithm, we initially set $\lambda = 10n^{-\frac{2}{9}}$  for the smoothing parameter. The smoothing parameter $\lambda$ is chosen by the generalized cross-validation (GCV) in the final step \cite{craven1978smoothing,gu1991minimizing}.  The iterative process is monitored for convergence by comparing two successive estimations, $\hat{\mu}_k$ and $\hat{\mu}_{k+1}$. Convergence is assumed when the norm of their difference satisfies the condition  $\| \hat{\mu}_{k+1} - \hat{\mu}_{k} \| \leq \epsilon, $ where $\epsilon$ is a predefined threshold and $\| \cdot \|$ is the $l^2$-norm.
%The final iteration step will separately choose a smoothing parameter via GCV process. to fit the final smoothing response. 
%During the iteration process, when the two successive estimation $\hat{\mu}_k$ and $\hat{\mu}_{k+1}$ satisfy $$\| \hat{\mu}_{k+1} - \hat{\mu}_{k} \| \leq \epsilon, $$
%with a given small threshold $\epsilon$ ($\| \cdot \|$ can be any vector norms, i.e., $\| \cdot \|_2$ in our package), we think the algorithm converges.
%At each iteration, The algorithm is considered to have converged when the $l^2$-norm of successive estimators of $\mu$ is sufficiently small. %Details are included in Appendix.
 %Different from the approach mentioned in \cite{ma2008penalized}, 
%See the supplementary materials for more details. %promising that $p$ is a smooth function. 
%For problem $(P_2)$, we apply the smoothing spline approach to estimate $\mu$ so that the estimation does not rely on the choice of knots and leads to a more robust result. 
%To solve problem $(P_2)$, we use smoothing spline fitting. 
%For given count data $\left\{y_i\right\}_{i=1}^n$ and corresponding pseudo-times $t_1,\cdots,t_n$, smoothing Spline for Poisson families uses the smoothness-penalized likelihood functional 
%\begin{equation}
%\label{pen}
%-\frac{1}{n} \sum_{i=1}^n (y_i\log \mu(t_i) - \mu(t_i) - \log(y_i!)) + \frac{\lambda}{2}J(\mu)\end{equation}
%where $\mu\in \mathcal{H}$.  The specified derivation is provided in the supplementary material.  
The proposed method is detailed in Algorithm 1.

\section{Simulation Studies}
This simulation study is designed to demonstrate the finite-sample performance of the proposed ZISS algorithm, specifically in terms of mean curve fitting and probability curve estimation. We compared ZISS with the following existing methods: (1) the direct smoothing spline (DSS) method, which fits smoothing spline ANOVA to all data points including zeros, (2) the non-zero fit (NZSS) method, which fits smoothing spline ANOVA to all non-zero data points, and (3) the ZIGAM method \cite{barry2002generalized}, fitting all data points. 

%Intuitively, the most direct way to fit the curve is to apply smoothing spline fitting to all of the data points, ignoring the patterns of zeros in the data. This approach is denoted as Direct Smoothing Spline (DSS) in the experiment analysis and will be proved inefficient. 

%Since the number of zeros in the data is expected to be inflated, another naive way to fit the model is removing all the zeros in the given data and then using the smoothing spline method. Such an approach is denoted as Non-Zero Fit (NZ) in our analysis and will also be proved under-behaved. 

% At the same time, we evaluated the ZIGAM Algorithm \cite{} to compare with our method in simulation studies. 
% We summarized these methods in table 1.

% \begin{center}
%     \begin{table}[H]
%         \centering
%         \begin{tabular}{lcccc}
%             \toprule
%             Method & Related Assumption  & Smoothness Control & return $p$ \\
%             \hline
%             ZISS(proposed) & No  & Continuous & Yes \\
%             NZ & No  & - & No \\
%             Direct-SSANOVA &No  & - & No \\
%             ZIGAM  & Yes  & Discrete & No \\
%             \hline 
%             \bottomrule
%         \end{tabular}
%         \caption{The Comparison of the Algorithms}
%     \end{table}
% \end{center}  
% Among the algorithms, Continuous Probability Zero-Inflated Poisson Model provides smooth control and return probability $p(t)$. It also does not require the smooth probability function $p$ relate to the response curve $\mu$, which is strongly required in ZIGAM Model.

\subsection{Simulation Settings}
We designed two simulation settings over the pseudotime interval $[0,1]$ to compare different methods. Our first experiment is designed to ensure that $p_1$ and $\mu_1$ are symmetric with respect to $t=0.5$, satisfying the condition which $p$ and $\mu$ are related, while the second experiment is designed with uncorrelated $p_2$ and $\mu_2$. In the second scenario, the data exhibit a higher incidence of zeros, making it challenging to accurately estimate the mean and zero probability functions.

For the first setting, the mean function is set as
$$\mu_1(t) = 2\sin(9t) + 2.5,$$
and the zero probability function is chosen as $$p_1(t) = \frac{1}{1+e^{-0.5(t-0.5)^2 + 1}},$$
where the pseudotime points $t_i,i=1,2,\cdots, N=41$ are  uniformly distributed within the time interval $[0,1]$. In this setting, $p_1$ and $\mu_1$ are both symmetric with respect to the central node $0.5$. There is a link function \cite{barry2002generalized, liu2008constrained}  $g$ to show the association between $\mu$ and $p$, such that $g(p) = \mu$,
$$\mu = 2\sin \left( 9\sqrt{2-2\log \left(\frac{1}{p}-1\right)} + 4.5\right) + 2.5.$$
At each pseudotime point $t_i$, we independently generated $M_i=M=80$ samples, $y_1,y_2,\cdots,y_M$ (corresponding to the respective pseudotime), using the zero-inflated Poisson model proposed in (\ref{model}). 

In the second experiment setting, we generated the data with the mean function $\mu(\cdot)$ according to
$$\mu_2(t) = \frac{8}{\sqrt{2\pi}} e^{-10(t-0.2)^2} + \frac{6}{\sqrt{2\pi}} e^{-100(t-0.7)^2},$$
and the true zero probability function 
 $$p_2(t) = \frac{1}{4} \sin (6t)+0.5.$$
In the second scenario, the mean function values within a specific region approximate zero, making them hard to distinguish from the true zeros in Poisson-distributed data. This presents a challenge for zero-inflated models that lack a flexible framework for the estimation of zero probability functions. Furthermore, the mean and zero probability functions are uncorrelated in this setting. We summarize the two settings in Table 1. 
\begin{center}
 \begin{table}[H]
 \label{simu}
 \centering
 \caption{Summary of the two simulation settings}
 \begin{tabular}
 { l|cc }

      \hline
     % \multicolumn{2}{|c|}{Notations} \\
     \hline
     \textbf{Curves} & \textbf{Setting 1} & \textbf{Setting 2} \\
     \hline
     $\mu(t)$ &$2\sin(9t) + 2.5$ &$\frac{8}{\sqrt{2\pi}} e^{-10(t-0.2)^2} + \frac{6}{\sqrt{2\pi}} e^{-100(t-0.7)^2}$ \\

     \hline
     $p(t)$ &$\frac{1}{1+e^{-0.5(t-0.5)^2 + 1}}$ & $\frac{1}{4} \sin (6t)+0.5.$\\
     \hline
     Correlation & Link function exists & Link function does not exist \\
     \hline
     Zero Inflation & Low & High\\
\hline
\hline
\end{tabular}
    \end{table}
    \end{center}
 %The important feature of this simulated data is that the true mean curve in $\mu_2$ has some values near 0. This means that there are more 'true' zeros (biological zeros) in the data compared with our setting 1.  At the same time, the setting indicates that $p_2$ and $\mu_2$ are highly unrelated, i.e., in this case, there are no link functions such that $g(\mu) = p$. 

   % \begin{figure}[H]
   % \begin{minipage}[t]{0.6\linewidth}
   % \centering
%    \includegraphics[height=4cm]{setting2.jpeg}
  %  \caption{Comparison of the 4 Methods Fitting Setting 2 \textcolor{blue}{(add legend, figure need update)}}
  %  \end{minipage}%
  %  \begin{minipage}[t]{0.6\linewidth}
   % \centering
   % \includegraphics[height=4cm,width=5cm]%{error2.jpeg}
  %  \caption{The Absolute Error of Estimated Probability Function}
  %  \end{minipage}
  %  \end{figure}

\begin{figure}[h!]
\label{response}
\centering
\includegraphics[height=12cm]{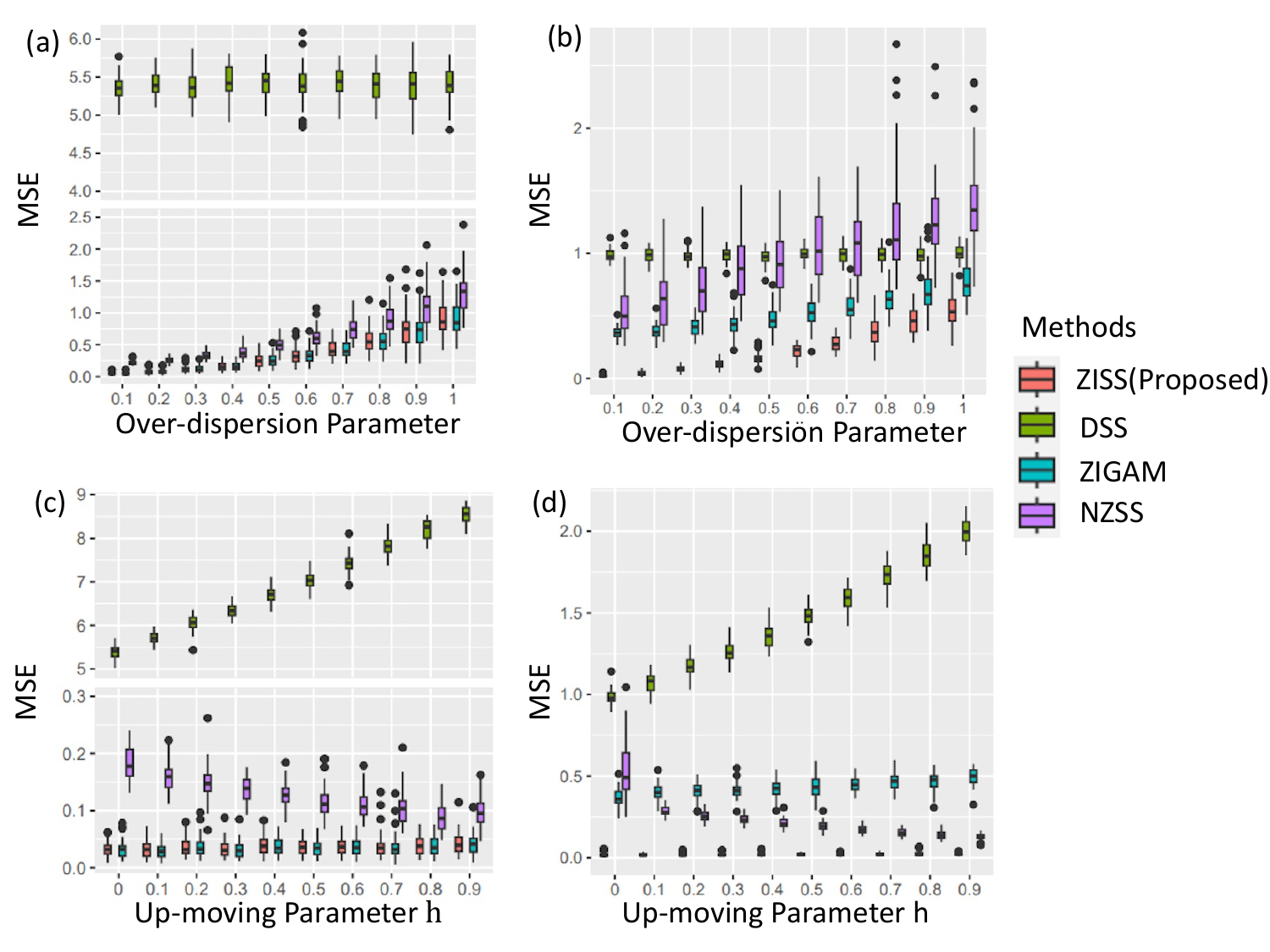}
\caption{Comparison of the mean curves estimated by different methods. In the figure, the $+$ symbol represents the observed data at each pseudotime point. Due to the possibility of multiple observations having identical values at a given pseudotime point, they might overlap in the figure. The black line illustrates the true mean curve $\mu_i$, while gray, orange, blue, and red lines represent the fitting results of DSS, NZSS, ZIGAM and ZISS, respectively. Below, the comparison between the estimated probability curve $\hat{p}$ and the true probability function $p$ is displayed. }

\end{figure}

To illustrate the numerical differences between these two simulation settings, we applied the methods for comparison to the simulated data from a single replication. In Figure 2, panels (a) and (c) depict the fitting results of Setting 1, while panels (b) and (d) present those of Setting 2. In Setting 1, ZISS and ZIGAM demonstrate comparable performances with accurate fitting patterns, attributed to the symmetry of true response curve $\mu$ and probability curve $p$. Conversely, DSS fitting underperforms, consistently falling below the true curve due to the inflated number of zeros in this setting. The NZSS method is effective  when $\mu(t)$ is significantly different from 0 but struggles when $\mu(t) \approx 0$. In Setting 2, the ZISS method offers the only estimate that aligns well with the true mean curve. ZIGAM appears over-smoothed, and DSS fitting remains below the true value, similar to its performance in Setting 1. The NZSS curve tends to overestimate the true curve. This setting's distinctive outcomes arise for several reasons: the zero inflation is predominantly due to zeros from two processes (i.e., Possion and zero processes) are not easily separable. 

%we find that only ZISS curve properly fits the true response. ZIGAM is over-smoothed while DSS fitting is below the true value as it is in setting 1. The NZ curve has a detectable delay and meanly performs higher than the true curve. Several reasons lead to the setting 2 performances: The zero-inflation is contributed more by the 'true' zeros (biological zeros), while the link function between $\mu$ and $p$ does not exist. Such pattern contributed to the poor performances of ZIGAM, DSS, and NZ.

To demonstrate the superiority of the ZISS method in both simulation settings, we conducted and reported the results from one hundred replicates. In each replicate, we generated new data according to the same mean and probability functions, and then fitted the data using the methods for comparison. We calculated the mean square error (MSE) between the estimated $\hat{\mu}$ and the true $\mu$ at each replicate and displayed the distributions of the MSEs for each method with their corresponding standard deviation (Std). The results of these repeated experiments are shown in Table 2. The results affirm the excellent performance of the ZISS method, characterized by the lowest MSEs coupled with the smallest Stds.

\begin{center}
    \begin{table}[H]
    \label{repeat}
        \centering
        \begin{tabular}{lcccc}
            \toprule
            Methods & ZISS (Proposed) & NZSS & ZIGAM & DSS  \\
            \hline
            MSE(Setting 1)  & {\bf 0.033} & 0.185 & {\bf 0.033} & 5.404 \\
            Std (Setting 1) & {\bf 0.012} & 0.033 & {\bf 0.012} & 0.165 \\
            MSE(Setting 2) &{\bf 0.169} & 0.550 & 0.368 & 0.971 \\
        Std (Setting 2) & {\bf 0.009} & 0.0198 & 0.051 & 0.052 \\
            \hline 
            \bottomrule
        \end{tabular}
        \caption{Comparison of mean square error (MSE) and standard deviation (Std) across different methods in 100 repeated experiments}
    \end{table}
\end{center}

\subsection{Study on the Impact of Zero-inflation}
In single-cell data, dropout (manifested as zero-inflation) and over-dispersion are two prevalent phenomena. In this section, we explore the impact of zero-inflation and over-dispersion, focusing on how various levels of zero-inflation and over-dispersion affect the performance of the proposed algorithm. To examine the increase in zero-inflation and over-dispersion, we utilized the simulation study as outlined in Ridout et al. (2001) \cite{ridout2001score}. Such a study serves as an effective method for analyzing the consequences of gradually diminishing signal intensity, which corresponds to an increase in zero-inflation. For the mean function $\mu$, we generated data that follows a negative binomial distribution with a mean of $\mu$ and a variance of $\mu(1+\sigma^2\mu)$. Here $\sigma^2=a$ is designated as the over-dispersion parameter. For each scenario and each over-dispersion parameter, we conducted 100 repetitions of the experiment and recorded the mean square error (MSE) for each method. To study the effect of reduced zero-inflation, we assessed the performance of the methods for comparison as the mean curve $\mu$ increasingly deviates from 0. For each specified up-moving parameter $h$, we conducted the experiments under both simulation settings, using data generated from $\mu +h$. These experiments were carried out with a progressively increasing average value of the mean curve. 

\begin{figure}[H]
\label{overdisp}
\centering
\includegraphics[width=16cm]{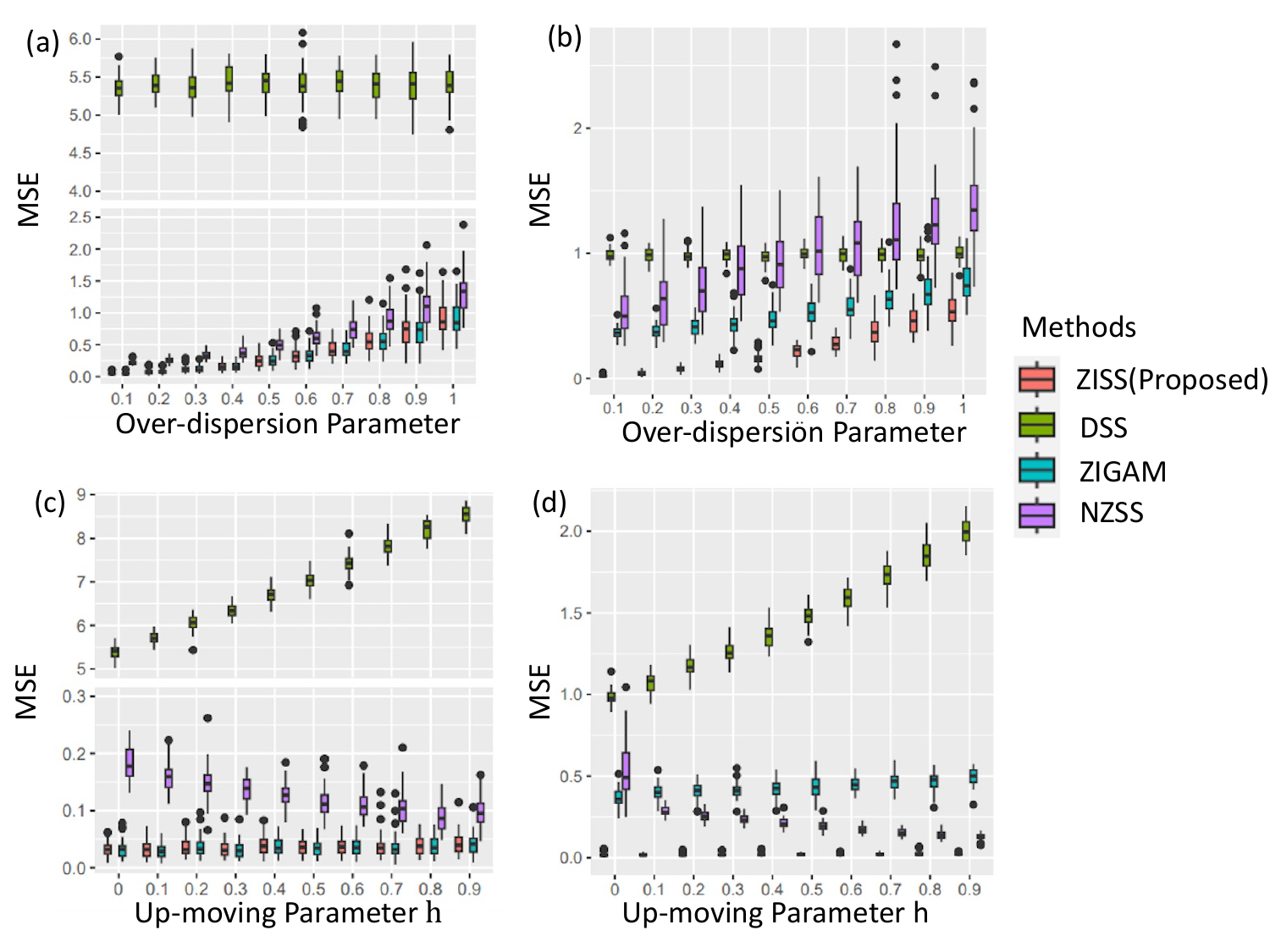}
\caption{Comparison of various methods across different over-dispersion and up-moving parameters in Setting 1 (panels a and c) and Setting 2 (panels b and d).}

\end{figure}

Figure 3 summarizes the results of over-dispersion and up-moving experiments for the two settings outlined in Section 3.1. Panels (a) and (b) illustrate the over-dispersion cases for settings $\mu_1$ and $\mu_2$, respectively. In addition, panels (c) and (d) show the MSE results of the methods for comparison for the up-moving experiments. For Setting 1, the results indicate that the ZIGAM and the ZISS methods perform similarly, whether the over-dispersion parameter or $h$ increases. Conversely, in Setting 2, the ZISS method consistently demonstrates superior performance in both over-dispersion and up-moving scenarios. This suggests that the proposed method is particularly effective in detecting and handling excessive zeros and over-dispersion within the data. 

Additionally, the boxplots reveal some notable trends. The performance of NZSS, a method that removes all zeros, tends to improve as the signal strength increases. However, this trend does not apply to the DSS method. Several factors contribute to these observations. Firstly, when the mean curve $\mu$ moves up, the separation between the zero and Poisson processes becomes more distinct. Consequently, in both settings, the performance of NZSS improves with an increase in the parameter $h$. Secondly, for the DSS fitting, when the mean function $\mu$ is modified to $\mu+h$, DSS directly fits the mean of $\mu+h$ and 0. This results in a larger discrepancy between the DSS fitted mean and the true mean curves as $h$ increases, leading to a rise in the MSE for DSS fitting. Lastly, the methods ZIGAM, NZSS, and ZISS tend to deteriorate as the over-dispersion parameter increases. This decline in performance can be attributed to the fact that increased over-dispersion leads to a larger variance in the data, causing more signals to be mistakenly identified as part of the zero process in the zero-inflated model.

%Several reasons contribute to this: when the curve $\mu$ is moving up, the biological zeros is indeed indicating that the performance of NZ will improve. As a result, one could see in both settings, the performances of NZ improved while increasing the up-moving parameter $h$ (defined in section 3.2). For the DSS fitting, when the response function $\mu(t)$ is replaced by $\mu(t)+h$, DSS will directly fit the mean of $\mu(t)+h$ and $0$. This means the larger the $h$, the larger the gap between DSS fitted response and the true response. Thus, one can see the increasing mean square error of DSS fitting. Finally, methods ZIGAM, NZ, and ZISS become worse when the over-dispersion parameter increases, since we 

\section{Real Data Analysis}
In this study, we conducted an analysis of scRNA-seq data obtained from the human kidney cell atlas. The researchers in this kidney disease study employed single-cell assays on diseased kidneys, providing valuable insights into the mechanisms underlying the increased risk of chronic kidney disease and kidney failure resulting from acute kidney injury events \cite{lake2021atlas}. We obtained the raw count matrix of integrated single-nucleus and scRNA-seq data of the adult human kidney, which was generated using the 10x Genomics platform. The raw data consists of 33,920 genes and 304,652 cells. To focus on the epithelial cell of the proximal tubule in scRNA-seq data, the data is reduced to 33,920 genes and 24,359 cells. To ensure data quality, cells with fewer than 500 reads and more than 50\% of reads mapping to the mitochondrial genome were identified as low quality and excluded from further analysis. Our focus in this study is on patients with acute kidney failure (n=12) and chronic kidney disease (n=15), resulting in a total of 19,507 cells from 27 patients being used for data analysis. Additionally, we utilized 122 kidney-disease related genes specifically identified in this kidney study\cite{lake2021atlas}. Considering that the data was collected from multiple laboratories, we applied the Harmony method to address any batch effects and ensure accurate analysis \cite{korsunsky2019fast}. We applied the cluster-based minimum spanning tree (cMST) method to estimate the pseudotime for each cell \cite{ji2016tscan, hou2021statistical}. Once the pseudotime was acquired, we partitioned the entire time interval (0-20,000) into 150 sub-intervals. Thus, cells were assigned to one of these sub-intervals based on their estimated pseudotime. In total, we obtained 150 pseudotime points for each gene of a participant. In this data analysis, we conducted a comparison between the proposed method and ZIGAM under various conditions. Our findings demonstrate that the proposed method consistently performs well across various levels of sparsity (i.e., proportion of zeros) and different numbers of observations.

%To focus on the epithelial cell of the proximal tubule in scRNA-seq data, the data is reduced to 33,920 genes and 24,360 cells.  

%We use our algorithm to process the data, find the probability curve $p(t)$ under our assumption, and fit the curve $\mu(t)$.
%Since the responses (values) are integers, we use our Zero-inflated Poisson Model. After deleting all the NA values, we round off the time points with an accuracy of 0.1 to ensure a single time point of the data contains various responses. 

\subsection{The performance of the proposed method remains consistent across various levels of sparsity.}
In this kidney study, a key research objective was to investigate differential gene expression patterns between participants with acute kidney failure and chronic kidney disease. To address this question, we selected two important genes (Ensembl IDs: ENSG00000137673 and ENSG00000118785) associated with chronic kidney disease and applied both the proposed method and ZIGAM to the data. The gene expression levels in the participant with acute kidney failure show a high expression across the pseudotime from 10,000 to 20,000. In contrast, the gene expression levels in the participant with chronic kidney disease exhibit a high expression throughout the pseudotime from 0 to 20,000. In Figure 4 (a), the gene expression data show high sparsity levels across the pseudotime range of 0 to 5,000. The proposed method demonstrates superior performance, capturing the underlying pattern accurately, while the ZIGAM method exhibits an undersmoothed fit. When the sparsity levels are low (e.g., pseudotime from 10,000 to 20,000), both methods perform comparably in capturing the gene expression dynamics.

\begin{figure}[h!]
\label{real1}
\centering
\includegraphics[height=6.5cm]{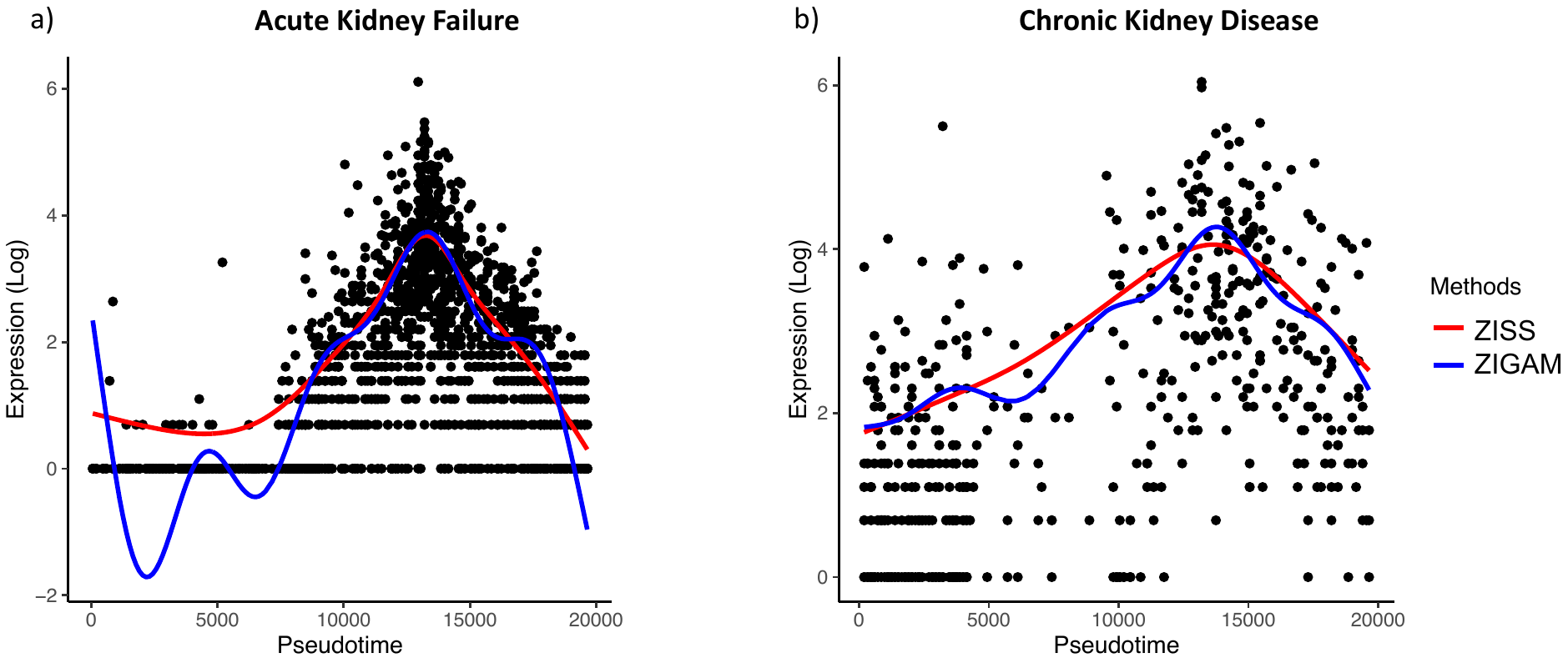}
\caption{The fitted curves for two subjects with acute kidney failure and chronic kidney disease using two different methods: ZISS (a) and ZIGAM (b). The curves represent the expression profiles of genes identified by their Ensembl IDs, with ENSG00000137673 used for panel (a) and ENSG00000118785 used for panel (b). }

\end{figure}

\subsection{The performance of the proposed method remains consistent for different numbers of observations.}
We utilized the IDEAS method \cite{zhang2022ideas}, a differential gene expression detection approach for individual-level single-cell data, to analyze the dataset, resulting in the identification of 13 differentially expressed genes out of the initial set of 122 genes. Among these genes, we selected the one with the highest total read count for further analysis. In Figure 5, we display the fitted curves generated by the proposed method and ZIGAM for this particular gene in four participants. As shown in Figure 5 (a) and (b), when a larger amount of data is available, both the proposed method and ZIGAM exhibit satisfactory performance. However, when the number of observations is limited, the proposed method outperforms ZIGAM, which tends to suffer from issues related to under-smoothing, see Figure 5 (c) and (d). The issue may stem from selecting an inappropriate smoothing parameter using GCV in ZIGAM due to the limited sample size.

\begin{figure}[h!]

\centering
\includegraphics[scale=0.48]{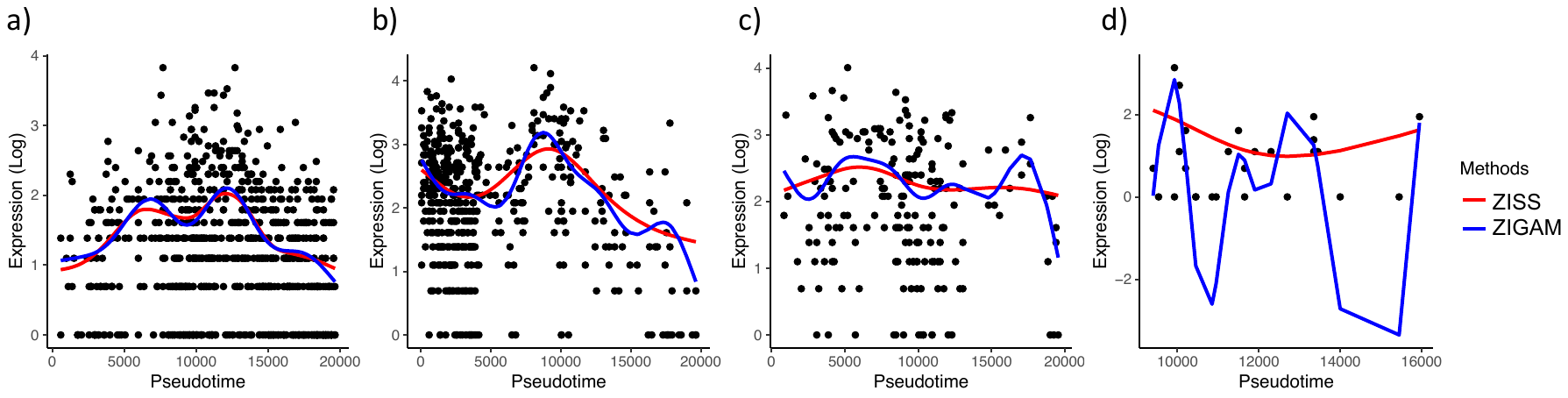}
\caption{The proposed method is robust to various data levels in the data of four participants. }

\end{figure}

%In the experiment, we compare the CPZIP and the SSANOVA fitting by directly removing 0 from the data(Non-Zero-Response). The graph shows the fitted and probability curves $p(t)$. 
\section{Conclusion}
In conclusion, our comprehensive analysis and experimental results strongly advocate for the effectiveness of the Zero-Inflated Smoothing Spline (ZISS) method in handling datasets characterized by zero inflation and over-dispersion. The ZISS method has demonstrated a robust capability in accurately identifying and interpreting signals within such complex datasets, particularly excelling in scenarios where traditional models struggle. 

In summary, the ZISS method exhibits exceptional performance in scenarios characterized by high zero-inflation rates, adeptly differentiating between true zeros and technical zeros in single-cell datasets. Additionally, this method demonstrates resilience in environments with various levels of over-dispersion, consistently maintaining both accuracy and reliability. Moreover, while specifically designed for single-cell temporal data, the ZISS method is versatile enough to be applied effectively in other domains featuring temporal and/or spatial data with excessive zeros.

%\section{Acknowledgements}
%The R package ZISS is available at xx.

\newpage
\bigskip
\begin{center}
{\large\bf Appendix}
\end{center}

%\begin{description}

%\item[Title:] Brief description. (file type)

%\item[R-package for ZISS:] \textcolor{red}{Yifu: I asked for another double check}

%\item[The data set:] Data set used in section 4.

%\end{description}
\appendix
\section{List of Notations}
\begin{tabular}{ |p{3cm}||p{11cm}|  }
%\centering
 % \hline
 % \multicolumn{2}{|c|}{Notations} \\
 \hline
 \textbf{Notations} & \textbf{Explanations} \\
 \hline
$t_i$ & the $i$th pseudotime point\\
 \hline
 $M_{i}$ & the number of observations at the pseudotime point $t_i$ \\
 \hline
 $y_{i,j}$ & the $j$-th observation at the $i$th pseudotime point\\
 \hline
 $g_{i,j}$ & the latent variable indicating zero processes \\
 \hline
 $p(t_i)$ & the probability function, denoting the probability from the Poisson process at the psutotime point $t_i$.\\
 \hline
 $\alpha_i$ & the coefficients of B-spline basis functions  \\
 \hline
 $b_i(\cdot)$ & the B-spline basis functions\\
 \hline
 $J(\cdot)$ & the penalized functional to measure the smoothness of functions, quadratic semi-norm in reproducing kernel Hilbert space \\
 \hline
 $\boldsymbol{\alpha}$ & a vector of coefficients $(\alpha_1,\cdots,\alpha_m)$\\
 \hline
 $\mu$ & a smooth function denoting the mean of the Poisson process\\
 \hline 
 $\lambda$ & smoothing parameter to control the smoothness of mean function estimate\\
 \hline
 $\hat{\mu}$ & the estimate of $\mu$ \\
  \hline
 $\hat{p}$ & the estimate of $p$\\
 \hline
 
\end{tabular}

\section{Derivation Details}
%\subsection{Details in Section 2.1}
%We elaborate the details in section 2.1 in this section. 
%\subsection{Details in Section 2.2}
%We elaborate on the derivation in section 2.2 in this section of supplementary material.
Given the single-cell temporal data $y_{i,j}$, the likelihood function $L$ is given as 
$$L(\mu,p,y) = \prod_{i=1}^N \prod_{j=1}^{M_i} \left( p(t_i)\frac{\mu(t_i)^{y_{i,j}}}{y_{i,j}!}\right)^{g_{i,j}} \left( 1-p(t_i)\right)^{1-g_{i,j}}.$$
We add a penalty on the smoothness of $\mu$ to the logarithm of $L$. Then the negative penalized log-likelihood function is
\begin{equation}
\begin{aligned}l(\mu,p,y) = &-\sum_{i=1}^N \sum_{j=1}^{M_i} \left(g_{i,j} \log p(t_i) + (1-g_{i,j}) \log(1-p(t_i)) \right) \\
&- \sum_{i=1}^N \sum_{j=1}^{M_i} \left( g_{i,j} y_{i,j} \log \mu(t_i) -g_{i,j} \mu(t_i) \right) + \frac{\lambda}{2} J(\mu) + Const,
\end{aligned}
\end{equation}
where $J(\cdot)$ is the smoothness penalty functional and $\lambda$ is the smoothing parameter.  We apply Expectation-Maximum (EM) algorithm to estimate the functions $\mu$ and $p$, where $\mu$ is estimated by the smoothing spline ANOVA model and $p$ is obtained by using the $B$-spline basis expansion.

\paragraph{The Expectation (E) step} 
We first show the proof of Proposition 1.

\textit{Proof of Proposition \ref{cond-exp}.}
Given $\hat{\mu}^{(m)}$ and $\hat{p}^{(m)}$ from the $m$th iteration, we compute the expectation of each latent variable $g_{i,j}$ given observed data $y_{i,j}$, denoted by $q_{i,j}^{(m)}$.
Using the Bayesian rule, we have
\begin{equation}
\begin{aligned}
& \mathbb{E}\left[ g_{i,j} | y_{i,j}=0,\hat{\mu},\hat{p} \right] = \mathbb{P}\left[ g_{i,j} = 1| y_{i,j}=0,\hat{\mu},\hat{p} \right] \\
&=\frac{\mathbb{P}\left[g_{i,j} = 1, y_{i,j}=0|\hat{\mu},\hat{p} \right]}{\mathbb{P}\left[ y_{i,j} =0|\hat{\mu},\hat{p} \right]} \\
&= \frac{\mathbb{P}\left[ y_{i,j}=0|g_{i,j} = 1,\hat{\mu},\hat{p} \right] \mathbb{P}[g_{i,j} =1|\hat{\mu},\hat{p}]}{\mathbb{P}\left[ y_{i,j}=0|g_{i,j} = 1,\hat{\mu},\hat{p} \right] \mathbb{P}[g_{i,j} =1|\hat{\mu},\hat{p}] + \mathbb{P}\left[ y_{i,j}=0|g_{i,j} = 0,\hat{\mu},\hat{p} \right] \mathbb{P}[g_{i,j} =0|\hat{\mu},\hat{p}]}\\
&= \frac{e^{-\hat{\mu}(t_i)} \hat{p}(t_i) }{e^{-\hat{\mu}(t_i)} \hat{p}(t_i) + 1-\hat{p}(t_i)}\\
\end{aligned}
\end{equation}
Otherwise if $y_{i,j} \neq 0$, this implies that $g_{i,j}$ has to be $1$, as a result, $$\mathbb{E}[g_{i,j}|y_{i,j} \neq 0,\hat{\mu}, \hat{p}] = 1.$$ Thus, we complete the proof. $\square$

\paragraph{The Maximization (M) step} Given conditional expectation of $g_{i,j}$'s, i.e.,   $q_{i,j}$, we minimize the negative log-likelihood function $l$. Given that $l$ can be represented as the sum of individual terms for $\mu$ and $p$, it is sufficient to minimize these two terms separately. Proposition 2 implies that we only need to separately optimize problems $P_1$ and $P_2$. For $P_1$, we have
$$(P_1)~~\min_{p\in \mathcal{P}} -\sum_{i=1}^N \sum_{j=1}^{M_i} \left(q_{i,j} \log p(t_i) + (1-q_{i,j}) \log(1-p(t_i)) \right),$$
where $\mathcal{P}$ is some function space we limit the estimate of $p$ into. We estimate the probability function by assuming $p(t)$ has the form
$$p(t) = \frac{1}{1+exp(\sum_{i=1}^m\alpha_i b_i(t))},$$
where $b_i(t)$'s are the B-spline basis functions and $\alpha_1,\cdots,\alpha_m$ are the corresponding coefficients. In this case, $\log \left( \frac{p(t)}{1-p(t)} \right) = -\sum_{i=1}^m \alpha_i b_i(t)$. The advantages of adopting this particular form are detailed in Section 2.2. Then in the M step, we need to estimate the parameters $\alpha_1,\cdots,\alpha_m \in \mathbb{R}$ to obtain the estimate of $p$. In particular, we aim to maximize
$$\sum_{i=1}^N \sum_{j=1}^{M_i} q_{i,j}\log p(t_i) + (1-q_{i,j}) \log (1-p(t_i)).$$
With the form of $p(t)$, we only need to maximize 
$$F(\alpha_1,\cdots,\alpha_m) = -\sum_{i=1}^N \sum_{j=1}^{M_i} q_{i,j}\sum_{k=1}^m \alpha_k b_k(t_i) + \sum_{i=1}^N \sum_{j=1}^{M_i} \log \frac{1}{1+exp(-\sum_{k=1}^m \alpha_k b_k(t_i))}.$$
This is a multi-variable function of $\boldsymbol{\alpha} = (\alpha_1,\cdots,\alpha_m)$, so we use the Newton's method \cite{traub1979convergence} to find the maximum of $F(\alpha_1,\cdots,\alpha_m)$ with the following updating rule,
$$\boldsymbol{\alpha} \leftarrow \boldsymbol{\alpha} - (\nabla_{\boldsymbol{\alpha}} ^2 F(\boldsymbol{\alpha}))^{-1} \nabla_{\boldsymbol{\alpha}} F(\boldsymbol{\alpha}).$$
The derivatives on $\alpha_k$, 
$$\frac{\partial F}{\partial \alpha_k} = -\sum_{i=1}^N \sum_{j=1}^{M_i}q_{i,j}b_k(t_i) + \sum_{i=1}^N \sum_{j=1}^{M_i} \frac{b_k(t_i)}{1+\exp (\sum_{l}\alpha_l b_l(t_i))},$$
constructs the vector $\nabla_{\boldsymbol{\alpha}}F(\boldsymbol{\alpha})$, while the second-order derivatives with respect to $\alpha_k$ and $\alpha_l$, 
$$\frac{\partial^2 F}{\partial \alpha_k \partial \alpha_l} = \sum_{i=1}^N \sum_{j=1}^{M_i}\frac{b_k(t_i)b_l(t_i) \exp(\sum_{r=1}^m\alpha_r b_r(t_i)))}{(1+\exp(\sum_{r=1}^m\alpha_r b_r(t_i)))^2},$$
constructs the Hessian of $F$ with respect to $\boldsymbol{\alpha},$ $\nabla^2_{\boldsymbol{\alpha}}F(\boldsymbol{\alpha})$.

We use the penalized likelihood method to solve the problem $P_2$, 
$$(P_2)~~\min_{\mu} - \sum_{i=1}^N \sum_{j=1}^{M_i} \left( q_{i,j} y_{i,j} \log \mu(t_i) -q_{i,j} \mu(t_i) \right) + \frac{\lambda}{2} J(\mu).$$
Based on the Representer Theorem, we incorporate the form of $\mu(t)$ as given in equation (2) into the objective function above and apply Newton's method for optimization. Readers may see \cite[Section 5]{gu2013smoothing} for more details of estimation. 

\newpage

\bibliographystyle{unsrt}

\bibliography{ref}
\end{document}